\documentclass[12pt]{article}
\usepackage{amsmath,amssymb,amsfonts,amsthm}

\title{Representations of Coherent and Squeezed States in an Extended Two-parameters Fock Space}

\author{M K Tavassoly$^{1, 2}$,  M H Lake$^{1}$
\\
\footnotesize{1- Department of Mathematical Sciences, Yazd University, Yazd, Iran} \\ \footnotesize{2- Research Group of Optics and Photonics,
Yazd University, Yazd, Iran}
\\ \footnotesize{E-mail: mktavassoly@yazduni.ac.ir;  }}

\begin{document}
\maketitle

\begin{abstract}
  Recently a $f$-deformed Fock space which is spanned by $|n\rangle_{\lambda}$ has been introduced. These bases are indeed the eigen-states of a deformed non-Hermitian
  Hamiltonian. In this contribution, we will use a rather new non-orthogonal  basis vectors for the construction of coherent and
  squeezed states, which in special case lead to the earlier known states. For this purpose, we first generalize the previously introduced Fock space
  spanned by $|n\rangle_{\lambda}$ bases, to a new one, spanned by an extended two-parameters bases $|n\rangle_{\lambda_{1},\lambda_{2}}$. These bases are now the
  eigen-states of a non-Hermitian Hamiltonian $H_{\lambda_{1},\lambda_{2}}=a^{\dagger}_{\lambda_{1},\lambda_{2}}a+\frac{1}{2}$,
  where $a^{\dagger}_{\lambda_{1},\lambda_{2}}=a^{\dagger}+\lambda_{1}a + \lambda_{2}$ and $a$ are respectively, the deformed
  creation and ordinary bosonic annihilation operators. The bases $|n\rangle_{\lambda_{1},\lambda_{2}}$ are non-orthogonal
  (squeezed states), but normalizable. Then, we deduce the new representations of coherent and squeezed states, in our two-parameters
  Fock space. Finally, we discuss the quantum statistical properties, as well as the non-classical properties of the obtained states, numerically.
\end{abstract}

{\bf Keywords:} Coherent State, Squeezed State, Representation Theory, Quantum Statistics, Non-orthogonal Bases.

{\bf Pacs:} 42.50.Dv, 42.50.Ar, 03.65.Fd

\newcommand{\I}{\mathbb{I}}
\newcommand{\norm}[1]{\left\Vert#1\right\Vert}
\newcommand{\abs}[1]{\left\vert#1\right\vert}
\newcommand{\set}[1]{\left\{#1\right\}}
\newcommand{\R}{\mathbb{R}}
\newcommand{\C}{\mathbb C}
\newcommand{\eps}{\varepsilon}
\newcommand{\To}{\longrightarrow}
\newcommand{\BX}{\mathbf{B}(X)}
\newcommand{\HH}{\mathfrak{H}}
\newcommand{\D}{\mathcal{D}}
\newcommand{\N}{\mathcal{N}}
\newcommand{\la}{\lambda}
\newcommand{\af}{a^{ }_F}
\newcommand{\afd}{a^\dag_F}
\newcommand{\afy}{a^{ }_{F^{-1}}}
\newcommand{\afdy}{a^\dag_{F^{-1}}}
\newcommand{\fn}{\phi^{ }_n}
 \newcommand{\HD}{\hat{\mathcal{H}}}
 \newcommand{\NDD}{\mathcal{N}}



\section{Introduction}
       Orthonormal set of bases, $\{|n\rangle , n\in N,   \langle m | n \rangle = \delta_{m, n}\}_{n=0}^{\infty}$,
       usually as the eigenvectors of hermitian Hamiltonians,
       is the most common set of bases has been
       frequently used in the framework of mathematical description of many areas of
       physics, especially in quantum mechanics. Moreover, the generalization to non-orthogonal basis vectors such as $|n\rangle_{\lambda}$
       proposed in \cite{drtvassoly1} may be preferred. The non-orthogonal basis states are recognized to be helpful in "generalized
       measurement", "quantum non-demolition measurement" and "quantum
       information theory" \cite{peres}. In the present paper, we will extend the previous work in \cite{drtvassoly1} and introduce a
       set of two-parameter non-orthogonal bases $\{|n\rangle_{\lambda_{1},\lambda_{2}}\}_{n=0}^{\infty}$ which are indeed the eigenvectors of a special
       deformed (non-hermitian) harmonic oscillator Hamiltonian. Also, following this idea will lead us to some new classes of generalized coherent states (CSs) and squeezed states  (SSs)
       in the quantum optics field. Generalized CSs   not only possess beautiful mathematical structure but also
       can generally be useful for the description of various concepts of physics \cite{vogel, chinees1, chinees2}
       (also see \cite{chinees-revised}).
       Along the latter goal, we will construct CSs  and SSs, as venerable objects in physics \cite{Ali2000}, using the bases  $\{|n\rangle_{\lambda_{1},\lambda_{2}} , n\in N\}_{n=0}^{\infty}$, instead of the usual orthonormal
       $\left\{|n\rangle\right\}_{n=0}^{\infty} $ and non-orthogonal $\left\{|n\rangle_{\lambda}\right\}_{n=0}^{\infty}$ bases in
       \cite{drtvassoly1}. Thus, in our opinion this approach provides a more fundamental and certainly more flexible
       basis, than that of the usual orthonormal one, considered extensively in the  literature, as well as the non-orthogonal
       basis $|n\rangle_{\lambda}$ in \cite{drtvassoly1}, for the construction of new generalized CSs and SSs \cite{Ali2004, klauder, perelomov}.
       Indeed, the explicit forms of CSs and SSs will be introduced, which contain two tunable parameters $\lambda_{1}$ and $\lambda_{2}$,
       by which one can obtain wide classes of states, from "standard CSs and SSs" (with $\lambda_{1}=0=\lambda_{2}$),
       previous non-orthogonal representations of CSs and SSs in \cite{drtvassoly1} (with $\lambda_{1}=0$) to new
       two-parameter states, will be introduced here. \\
       For a quantized harmonic oscillator (QHO), one has $H_{0}= a^{\dagger}a+\frac{1}{2}$, $[a,a^{\dagger}]=\hat{I}$,
       $(a^{\dagger})^{\dagger}=a$, where $a$, $a^{\dagger}$, $H$  and $\hat{I}$ are respectively bosonic annihilation,
       creation, Hamiltonian and unity operators. We consider the parametric harmonic oscillator, by deforming the creation
       operator according to the proposal in \cite{xiang}
\begin{eqnarray}\label{mh02}
     a^{\dagger}_{\lambda_{1},\lambda_{2}}=a^{\dagger}+\lambda_{1}a+\lambda_{2}\hat I, \qquad \lambda_{1},\lambda_{2} \in R
\end{eqnarray}
     while the annihilation operator remains unchanged, i.e., $ a_{\lambda_{1},\lambda_{2}}=a$. So,
     by analogy with the harmonic oscillator, the $\lambda_{1},\lambda_{2}$-Hamiltonian becomes
\begin{eqnarray}\label{mh03}
     H_{\lambda_{1},\lambda_{2}}&=&\frac{1}{2}\lbrace a^{\dagger}_{\lambda_{1},\lambda_{2}},a \rbrace \quad \nonumber\\
     &=&a^{\dagger}a+\lambda_{1}a^{2}+\lambda_{2}a+\frac{1}{2}.
\end{eqnarray}
     Note that,
     $(a^{\dagger}_{\lambda_{1},\lambda_{2}})^{\dagger} \neq a_{\lambda_{1},\lambda_{2}}$ and $H^{\dagger}_{\lambda_{1},
     \lambda_{2}} \neq H_{\lambda_{1},\lambda_{2}}$. But, as for QHO yet we have
\begin{eqnarray}\label{mh04}
     [a,a^{\dagger}_{\lambda_{1},\lambda_{2}}] &=& \hat{I}, \quad
     [H_{\lambda_{1},\lambda_{2}},a^{\dagger}_{\lambda_{1},\lambda_{2}}]=a^{\dagger}_{\lambda_{1},\lambda_{2}},\nonumber \\
     && [H_{\lambda_{1},\lambda_{2}},a]=-a.
\end{eqnarray}
       Thus, the new set $\left\{a,a^{\dagger}_{\lambda_{1},\lambda_{2}},a^{\dagger}_{\lambda_{1},\lambda_{2}} a,
       \hat{I}\right\}$ still satisfies the Weyl-Heisenberg algebra as in the case of QHO.\\
       One of us has shown in  \cite{Ali2004} that, a large class of generalized CSs can be obtained by changing the
       bases in the underlying Hilbert space. At this stage, we would like to illustrate that, the particular deformation which we
       employed in this paper is a special case of the general scheme for the representation theory of CSs,  has been
       introduced in \cite{Ali2004}. To clarify more, we will explain briefly the setting. Let $\mathcal{H}$ be a
       Hilbert space and $T,T^{-1}$ be operators densely defined and closed on $\mathcal{D}(T)$ and $\mathcal{D}(T^{-1})$,
       respectively, and $F=T^{\dagger}T$. Two new Hilbert spaces $\mathcal{H}_{F}$, $\mathcal{H}_{F^{-1}}$ are the
       completions of the sets  $\mathcal{D}(T)$ and $\mathcal{D}({T^{\dagger}}^{-1})$ with the scalar product
       $\langle f|g\rangle_{F}=\langle f|Fg\rangle_{\mathcal{H}}$ and  $\langle f|g\rangle_{F^{-1}}=\langle f|F^{-1}g\rangle_{\mathcal{H}}$,
       respectively. Considering the generators of the Weyl-Heisenberg algebra, as basis on $\mathcal{H}$, one may obtain
       the transformed generators on $\mathcal{H}_{F}$,  as
       follows
\begin{eqnarray}\label{mh106}
    a_{F}=T^{-1}a T, \qquad a^{\dagger}_{F}=T^{-1}a^{\dagger}T, \qquad
    N_{F}=T^{-1}NT.
\end{eqnarray}
     A similar argument may be followed for Hilbert space $\mathcal{H}_{F^{-1}}$.
     If we take the non-unitary $T$-operator as
\begin{eqnarray}\label{mh107}
      T_{n,\lambda_{1},\lambda_{2}}=\xi_{n}e^{-\frac{1}{2}\lambda_{1}a^{2}-\lambda_{2}a}
\end{eqnarray}
      it is easy to check that the following deformed operators may be obtained
\begin{eqnarray}\label{mh108}
      a_{F}=T_{n,\lambda_{1},\lambda_{2}}^{-1}a T_{n,\lambda_{1},\lambda_{2}}=a
\end{eqnarray}
\begin{eqnarray}\label{mh109}
      a^{\dagger}_{F}&=&T_{n,\lambda_{1},\lambda_{2}}^{-1}a^{\dagger}T_{n,\lambda_{1},\lambda_{2}} \nonumber\\
      &=&e^{\frac{1}{2}\lambda_{1}a^{2}\lambda_{2}a}a^{\dagger}e^{-\frac{1}{2}\lambda_{1}a^{2}-\lambda_{2}a}\nonumber\\
      &=&a^{\dagger}+\lambda_{1}a+\lambda_{2}\hat I
\end{eqnarray}
     and
\begin{eqnarray}\label{mh109a}
      N_{F}&=&T_{n,\lambda_{1},\lambda_{2}}^{-1}a^{\dagger}a T_{n,\lambda_{1},\lambda_{2}} \nonumber\\
      &=& e^{\frac{1}{2}\lambda_{1}a^{2}+\lambda_{2}a}a^{\dagger}a\; e^{-\frac{1}{2}\lambda_{1}a^{2}-\lambda_{2}a}\nonumber\\
      &=&a^\dag_{\lambda_{1},\lambda_{2}}a
 \end{eqnarray}
     where $\xi_{n}$ is an appropriate normalization factor, which will be determined in the continuation. In this
     manner,
     we have established the fundamental place of the particular kind of deformation proposed in \cite{xiang} and will
     be used by us in the present work, in the general framework of representation theory of CSs in non-orthogonal basis.\\
     In terms of $T_{n,\lambda_{1},\lambda_{2}}$, we can rewrite the Hamiltonian in  Eq. (\ref{mh03}), as
     $H_{\lambda_{1},\lambda_{2}}=T^{-1}_{n, \lambda_{1},\lambda_{2}}H_{0}T_{n, \lambda_{1},\lambda_{2}}$, where
     $H_{0}$ is the QHO Hamiltonian. According to the general formalism in \cite{Ali2004}, the normalized non-orthogonal
     bases in the new Hilbert space can be obtained as
\begin{eqnarray}\label{mh100}
     \vert n\rangle_{\lambda_{1},\lambda_{2}} &=& T_{n,\lambda_{1},\lambda_{2}}\vert n\rangle =\xi_{n}\sum_{r=0}^{n}\sum_{k=0}^{[\frac{n-r}{2}]} \nonumber\\
     &\times& \left(\frac{1}{2}\right)^{k}\frac{\lambda_{1}^{k}
     \lambda_{2}^{n-r-2k}}{k!(n-r-2k)!}\sqrt{\frac{n!}{r!}}\vert
     r\rangle,
\end{eqnarray}
     where we have used $T_{n,\lambda_{1},\lambda_{2}}$ in (\ref{mh107}), [m]
     denotes the integer part of $m$ and $\xi_{n}$ may be derived with
     the normalization  condition,
     $_{\lambda_{1},\lambda_{2}}\langle n|n\rangle_{\lambda_{1},\lambda_{2}}=1$ as
\begin{eqnarray}\label{mh07}
      \xi_{n} &=& [ \sum_{r=0}^{n} \frac{n!}{r!}\lambda_{2}^{2(n-r)}\sum_{k=0}^{[\frac{n-r}{2}]}\sum_{k'=0}^{[\frac{n-r}{2}]}(\frac{1}{2})^
     {k+k'}\nonumber \\ &\times&
     \frac{\lambda_{1}^{k+k'}\lambda_{2}^{-2(k+k')}}{k!(n-r-2k)!k'!(n-r-2k')!}] ^{-\frac{1}{2}}.
\end{eqnarray}
     Eq. (\ref{mh100}) suggests that, every state $|n \rangle_{\lambda_{1},\lambda_{2}}$ in this non-orthogonal
     Hilbert space can be regarded as a special finite superposition of $|0\rangle , |1\rangle ,...,|n\rangle$ in the standard Fock space.
     In the limit $\lambda_{1}\rightarrow 0$, $\xi_{n}$ reduces to the normalization factor
     $\left[L_{n}^{0}(-\lambda^{2})\right]^{-\frac{1}{2}}$ in one-parameter states ($\lambda_1=\lambda=\lambda_1$) introduced
     in \cite{drtvassoly1} ($L_{n}^{0}(x)$ are the Laguerre polynomials of $n$ the order) and
     in the limit $\lambda_{1},\lambda_{2}\rightarrow 0$ it reduces to unity.

     The non-Hermitian Hamiltonian $H_{\lambda_{1},\lambda_{2}}$, is isospectral with $H_{0}$ \cite{xiang},
     i.e.,
\begin{eqnarray}\label{mh08}
     H_{\lambda_{1},\lambda_{2}}|n\rangle_{\lambda_{1},\lambda_{2}}=E_{n,\lambda_{1},
     \lambda_{2}}|n\rangle_{\lambda_{1},\lambda_{2}},
\end{eqnarray}
     where $E_{n,\lambda_{1},\lambda_{2}}=E_{n}=n+\frac{1}{2}$ and $n=0,1,2,...$ . Also the actions of $a$ and
     $a^{\dagger}_{\lambda_{1},\lambda_{2}}$ on $\lambda_{1},\lambda_{2}$-bases are as follows
\begin{eqnarray}\label{mh10}
     a\vert n\rangle_{\lambda_{1},\lambda_{2}}&=&\frac{\xi_{n}}{\xi_{n-1}}\sqrt{n}\vert n-1\rangle_{\lambda_{1},\lambda_{2}}
\end{eqnarray}
     \begin{eqnarray}\label{mh13}
     a^{\dagger}_{{\lambda_{1},\lambda_{2}}}\vert n\rangle_{\lambda_{1},\lambda_{2}}&=&\frac{\xi_{n}}{\xi_{n+1}}
     \sqrt{n+1}\vert n+1\rangle_{\lambda_{1},\lambda_{2}}.
\end{eqnarray}
     As it may be expected, in the limit $\lambda_{1}\rightarrow 0$, we recover the results of the $\lambda$-states in
     \cite{drtvassoly1} and in the limit  $\lambda_{1},\lambda_{2}\rightarrow 0$, we get the usual actions of ladder
     operators of harmonic oscillator. The inner product of the states in (\ref{mh100}), also read as
\begin{eqnarray}
\label{mh12}
_{\lambda_{1},\lambda_{2}}\langle m \vert
     n\rangle_{\lambda_{1},\lambda_{2}} &=&
     \sqrt{m!n!} \; \xi_{n}\xi_{m}\sum_{r=0}^{min(m,n)}\frac{\lambda_{2}^{m+n-2r}}{r!}
     \sum_{k=0}^{[\frac{m-r}{2}]}\sum_{j=0}^{[\frac{n-r}{2}]}\left(\frac{1}{2}\right)^{k+j}\\ \nonumber
     &\times&
     \frac{\lambda_{1}^{k+j}\lambda_{2}^{-2(k+j)}}{k!(m-r-2k)!j!(n-r-2j)!}.
\end{eqnarray}

     Let us end this section, by mentioning some of the interesting and important points we may further conclude.
     \\
     a) It is important for our further work to establish that the non-orthogonal states $|n\rangle_{\lambda_{1},\lambda_{2}}$ can
     be regarded as the bases of new Fock space. For this purpose, the necessary and sufficient conditions mentioned
     for a one-dimensional quantum Fock space in \cite{bardek} will be investigated. In our case, these conditions are briefly,
    (i) existence of a vacuum state, such that $a|0\rangle=0$, note that we have considered $|0 \rangle_{\lambda_{1},
    \lambda_{2}}=|0\rangle$, (ii) $\langle 0|aa^{\dagger}_{\lambda_{1},\lambda_{2}}|0\rangle >0$, (iii) $[aa^{\dagger}_
    {\lambda_{1},\lambda_{2}},a^{\dagger}_{\lambda_{1},\lambda_{2}}a]= 0$ (in \cite{drtvassoly1} there's a typing error)
     and $aa^{\dagger}_{\lambda_{1},\lambda_{2}} \neq a^{\dagger}_{\lambda_{1},\lambda_{2}}a$.
    \\
    b) $a^{\dagger}_{\lambda_{1},\lambda_{2}}a |n\rangle_{\lambda_{1},\lambda_{2}}\equiv N_{\lambda_{1},\lambda_{2}}
    |n\rangle_{\lambda_{1},\lambda_{2}}=n|n\rangle_{\lambda_{1},\lambda_{2}}$,
    so $\hat{N}_{\lambda_{1},\lambda_{2}}$ can be regarded as number operator in the
    new Fock space. Moreover, from this equation, we see that $({n}+\lambda_{1}a^{2}+\lambda_{2}a)|n\rangle_
    {\lambda_{1},\lambda_{2}}= n|n\rangle_{\lambda_{1},\lambda_{2}}$, which indicates simply the ladder operator formalism
     \cite{wang2000} of the state $|n\rangle_{\lambda_{1},\lambda_{2}}$.\\
    \\
    c) By Eqs. (\ref{mh10}) and (\ref{mh13}), we obtain respectively
\begin{equation}\label{mh13a}
    a^{n}\vert n\rangle_{\lambda_{1},\lambda_{2}}=\xi_{n}\sqrt{n!}\vert 0 \rangle_{\lambda_{1},\lambda_{2}},
\end{equation}
\begin{equation}\label{mh14}
    \vert n\rangle_{\lambda_{1},\lambda_{2}}=\frac{\xi_{n}}{\sqrt{n!}}(a^{\dagger}_{{\lambda_{1},\lambda_{2}}})^{n}\vert 0
     \rangle_{\lambda_{1},\lambda_{2}},
\end{equation}
 which will be helpful for our next calculations.

    \section{Construction of CS$s$ in $|n\rangle_{\lambda_{1},\lambda_{2}}$ basis}\label{mhl03}

    Using the algebraic definition of CSs as the eigen-states of annihilation operator \cite{Das2002, manko1996}
\begin{eqnarray}\label{mh16}
    a\vert \alpha , {\lambda_{1},\lambda_{2}}\rangle=\alpha \vert \alpha , {\lambda_{1},\lambda_{2}}\rangle
\end{eqnarray}
    we want to construct the CSs in the new deformed bases introduced in Eq. (\ref{mh100}). As usual, expanding $\vert \alpha,
    {\lambda_{1},\lambda_{2}}\rangle$ in terms of $|n\rangle_{\lambda_{1},\lambda_{2}}$
    bases as $\vert \alpha , {\lambda_{1},\lambda_{2}}\rangle=\sum_{n=0}^{\infty}C_{n}\vert n\rangle_{\lambda_{1},\lambda_{2}}$,
    setting in Eq. (\ref{mh16}) and using Eq. (\ref{mh10}), we finally get
\begin{eqnarray}\label{mh18}
     \sum_{n=0}^{\infty}C_{n}\frac{\xi_{n}}{\xi_{n-1}}\sqrt{n} \vert n -1 \rangle_{\lambda_{1},\lambda_{2}} =
     \alpha\sum_{n=0}^{\infty}C_{n}\vert n \rangle_{\lambda_{1},\lambda_{2}}.
\end{eqnarray}
    The coefficients $C_{n}$ will be obtained as
\begin{eqnarray}\label{mh19}
    C_{n}=C_{0}\frac{\alpha^{n}}{\sqrt{n!}\xi_{n}}
\end{eqnarray}
    where we have used the fact that $\xi_{0}=1$. For the normalization condition of the state $\vert \alpha , {\lambda_{1},
    \lambda_{2}}\rangle$ one yields
\begin{eqnarray}\label{mh20}
    C_0=\exp\left[-\frac{1}{2}\lambda_{1}\Re(\alpha^{2})-\lambda_{2}\Re(\alpha)-\frac{\vert \alpha \vert ^{2}}{2}\right]
\end{eqnarray}
    where $\Re(\alpha)$ is the real part of $\alpha$. Finally, the normalized $\lambda_{1},\lambda_{2}$-CS takes the form
\begin{equation}
\label{mh21}
   \vert \alpha , {\lambda_{1},\lambda_{2}}\rangle = \exp\left[-\frac{1}{2}\lambda_{1}\Re(\alpha^{2})-\lambda_{2}
   \Re(\alpha)-\frac{\vert \alpha \vert ^{2}}{2}\right]
   \sum_{n=0}^{\infty}\frac{\alpha^{n}}{\sqrt{n!}\xi_{n}}\vert
   n\rangle_{\lambda_{1},\lambda_{2}},
\end{equation}
    which is a "\textit{new representation}" of CSs in $\vert n\rangle_{\lambda_{1},\lambda_{2}}$ basis.
    Their inner product, which allows over-completeness, can be expressed as
 \begin{eqnarray}
\label{mh22}
   \langle \alpha , \lambda_{1},\lambda_{2}|\beta , \lambda_{1},\lambda_{2}\rangle &=& N_{\alpha, \beta,
   \lambda_{1},\lambda_{2}}\sum_{m=0}^{\infty}\sum_{n=0}^{\infty}\alpha^{*m}\alpha^{n}
   \sum_{r=0}^{min(m,n)}\frac{\lambda_{2}^{m+n-2r}}{r!}\\ \nonumber && \sum_{k=0}^{[\frac{m-r}{2}]}
   \sum_{j=0}^{[\frac{n-r}{2}]}(\frac{1}{2})^{k+j}
   \frac{\lambda_{1}^{k+j}\lambda_{2}^{-2(k+j)}}{k!(m-r-2k)!j!(n-r-2j)!}
\end{eqnarray}
    with
\begin{equation}\label{mh23}
    N_{\alpha , \beta , \lambda_{1},\lambda_{2}}=\exp\left[-\frac{1}{2}\lambda_{1}(\Re(\alpha^{2})+\Re(\beta^{2}))
    -\lambda_{2}(\Re(\alpha)+\Re(\beta))-\frac{\vert \alpha \vert ^{2}}{2}-\frac{\vert \beta \vert ^{2}}{2}\right].
\end{equation}
    Now, we imply that, by Eq. (\ref{mh14}) the CSs in (\ref{mh21}) can be expressed in terms of the lowest
    eigen-state of $H_{\lambda_{1},\lambda_{2}}$ as
\begin{eqnarray}\label{mh24}
   \vert \alpha , {\lambda_{1},\lambda_{2}}\rangle=\exp\left[-\frac{1}{2}\lambda_{1}\Re(\alpha^{2})-\lambda_{2}
   \Re(\alpha)-\frac{\vert \alpha \vert ^{2}}{2}\right]e^{\alpha a^{\dagger}_{\lambda_{1},\lambda_{2}}}|0\rangle.
\end{eqnarray}
    By BCH lemma, (\ref{mh24}) can be rewritten as
%
\begin{eqnarray}
\label{mh25}
    \vert \alpha , {\lambda_{1},\lambda_{2}}\rangle &=& C_{0}e^{\frac{1}{2}\lambda_{1}\alpha^{2}+\lambda_{2}
    \alpha+\frac{\vert \alpha \vert ^{2}}{2}}\vert \alpha\rangle
    = e^{-\frac{1}{2}\lambda_{1}\Re(\alpha^{2})-\lambda_{2}\Re(\alpha)-\frac{\vert \alpha \vert ^{2}}{2}}\\ \nonumber &\times&
    e^{\frac{1}{2}\lambda_{1}\left(\Re(\alpha^{2})+i\Im(\alpha^{2})\right)+\lambda_{2} \left
    (\Re(\alpha)+i\Im(\alpha)\right)+\frac{\vert \alpha \vert ^{2}}{2}}\vert\alpha\rangle
    = D_{\lambda_{1},\lambda_{2}}(\alpha)\vert 0\rangle,
\end{eqnarray}

     where we assumed that  $D_{\lambda_{1},\lambda_{2}}(\alpha)=e^{i(\frac{1}{2}\lambda_{1}\Im(\alpha^{2})+\lambda_{2}\Im(\alpha))}D(\alpha)$
     and the imaginary part of $x$ is denoted by $\Im(x)$. Since $D(\alpha)\vert 0 \rangle = |\alpha\rangle $, which is the standard CSs,
     we conclude that $\vert \alpha , {\lambda_{1},\lambda_{2}}\rangle$ is identical to $|\alpha\rangle$, up to a phase factor and therefore
     $\vert \alpha , {\lambda_{1},\lambda_{2}}\rangle=|\alpha\rangle$ whenever $\alpha \in R$; a result that may be expected from the
     eigenvalue equation (\ref{mh16}). So, obviously there is no problem with resolution of the identity
\begin{eqnarray}\label{mh27}
    \int_{C} d\mu (\alpha) \vert \alpha , \lambda_{1} , \lambda_{2}\rangle \langle\alpha , \lambda_{1} , \lambda_{2}\vert = \hat{I}
\end{eqnarray}
    with $d\mu (\alpha)  \doteq \frac{1}{\pi} d^{2}\alpha$ .\\
    We would like to emphasize that, all we have done in this section is obtaining the explicit form of canonical CSs in a deformed Fock
    space $|n\rangle_{\lambda_{1},\lambda_{2}}$, which are non-orthogonal, and we called them new representation of canonical CSs.
    As another result, we conclude that by a particular superposition of $|n\rangle_{\lambda_{1},\lambda_{2}}$ bases (which
    themselves exhibit squeezing \cite{xiang}), we have obtained canonical CSs.
    Note that, both the orthogonal bases ${|n\rangle}$ which commonly have been used in the construction of CSs, and (one and two parameters)
    non-orthogonal bases introduced (and applied) by us, are
    non-classical states. But, orthogonal bases have sub-Poissonian statistics (due to the fact that their Mandel parameters are equal to $-1$)
    without squeezing, while non-orthogonal bases exhibit squeezing.

     The dynamical evolution of the $\lambda_{1},\lambda_{2}$-CSs in $|n\rangle_{\lambda_{1},\lambda_{2}}$ basis, may be simply
     obtained due to the linear spectrum feature of $H_{\lambda_{1},\lambda_{2}}$

\begin{eqnarray}
\label{m28}
    U(t)\vert \alpha , \lambda_{1} , \lambda_{2}\rangle &=& \exp\left[-\frac{1}{2}\lambda_{1}\Re(\alpha^{2})-
    \lambda_{2}\Re(\alpha)-\frac{\vert \alpha \vert ^{2}}{2}\right]\\ \nonumber &&
    \sum_{n=0}^{\infty}\frac{\alpha^{n}}{\sqrt{n!}\xi_{n}}e^{-itH_{{\lambda_{1},\lambda_{2}}}}\vert n\rangle_
    {\lambda_{1},\lambda_{2}}
    =e^{-\frac{it}{2}}\vert \alpha (t) , \lambda_{1} , \lambda_{2}\rangle
\end{eqnarray}
      where we have used $\alpha (t)\equiv\alpha e^{-it}$. This means that, the time evolution of $\lambda_{1},\lambda_{2}$-CSs
      in this non-orthogonal bases remains coherent in the same bases for all time (\textit{temporal stability}).

 \section{Construction of SS$s$ in $\lambda_{1},\lambda_{2}$ basis}\label{mhl05}
       According to the statement of Solomon and Katriel \cite{solomon}, the conventional SSs are obtained by the action of a
       linear combination of creation and annihilation operators on an arbitrary state. Now, by generalizing this procedure to
       $a^{\dagger}_{\lambda_{1},\lambda_{2}}$ introduced in (\ref{mh02}) and $a_{\lambda_{1},\lambda_{2}}=a$ of the deformed oscillator we have
 \begin{eqnarray}\label{mh31}
      (a-\eta a^{\dagger}_{\lambda_{1},\lambda_{2}})|\eta,\lambda_{1},\lambda_{2}\rangle=0\qquad \eta \in C.
 \end{eqnarray}
      This equation for $\lambda_{1},\lambda_{2}=0$  leads to the squeezed vacuum states
      $|\eta \rangle= C_{0}\exp\left[\frac{\eta (a^{\dagger})^{2}}{2}\right]|0\rangle$, where $C_{0}$ is a suitable normalization
      coefficient. But, in general, from Eq. (\ref{mh31}) and by similar procedure we have done in previous section
      for $\lambda_{1},\lambda_{2}\neq 0$, we will arrive at a "\textit{new representation}" for $\lambda_{1},\lambda_{2}$-SSs as
 \begin{eqnarray}\label{mh34}
      \vert \eta,\lambda_{1} , \lambda_{2}\rangle =C_{0} \sum_{n=0}^{\infty}\frac{\eta^{n}}{\xi_{2n}}
      \sqrt{\frac{(2n-1)!!}{(2n)!!}}\vert 2n \rangle_{\lambda_{1} , \lambda_{2}}
 \end{eqnarray}
      where $\xi_{2n}$ may be obtained from Eq. (\ref{mh07}).
      For the normalization factor, one may get
\begin{eqnarray}\label{m28}
C_{0} &=& [\sum_{n=0}^{\infty} \sum_{m=0}^{\infty} \eta^{n} \eta^{*m}(2n-1)!!(2m-1)!! \sum_{r=0}^{min(2m , 2n)}
   \frac{\lambda_{2}^{2m+2n-2r}}{r!}\\ \nonumber &&
   \sum_{k=0}^{[\frac{2m-r}{2}]} \sum_{j=0}^{[\frac{2n-r}{2}]} \left( \frac{1}{2}\right)^{k+j}
   \frac{\lambda_{1}^{k+j}\lambda_{2}^{-2(k+j)}}
   {k!(2m-r-2k)!j!(2n-r-2j)!}]^{-\frac{1}{2}}.
\end{eqnarray}
      These $\lambda_{1},\lambda_{2}$-SSs are normalizable, provided that the coefficients $C_{0}$ is nonzero and finite.
      From the above discussion, we immediately conclude that translating $|\eta , \lambda_{1},\lambda_{2}\rangle$
      (in $|n\rangle_{\lambda_{1},\lambda_{2}} $ basis) to a state in the standard Fock space $|n\rangle$ does
      not coincide with the one we previously referred to in the beginning of this section. This is due to the fact
      that, unlike the introduced CSs in Eq. (\ref{mh21}), in obtaining the  $\lambda_{1},\lambda_{2}$-SSs both the annihilation and the deformed
      creation operators are contributed (see Eq. (\ref{mh31})).

\section{Non-classical properties}\label{mhl06}

       In this section, we will introduce the non-classical criteria which we will consider in our numerical results.

     \subsection{Mandel parameter }\label{mhl07}

       The standard CSs possess the Poissonian  distribution as
       $ P(n)=|\langle n|\alpha\rangle|^{2}= e^{-|\alpha|^{2}}{|\alpha|^{2n}}/{n!}$, whose mean and variance
        are equal to $|\alpha|^{2}$. Similarly, in the case of our $\lambda_{1} , \lambda_{2}$-CSs we can define:
\begin{eqnarray}\label{mh37}
      P_{\lambda_{1} , \lambda_{2}}(n) = \vert _{\lambda_{1} , \lambda_{2}}\langle n \vert \alpha , \lambda_{1} , \lambda_{2}\rangle
      \vert^{2},
\end{eqnarray}
      which may be interpreted as the probability of finding the states $|\alpha,\lambda_{1},\lambda_{2}\rangle$
      in non-orthogonal $|n\rangle_{\lambda_{1},\lambda_{2}}$ basis.
      So,
 \begin{eqnarray}\label{mh38}
      _{\lambda_{1} , \lambda_{2}}\langle n \rangle_{\lambda_{1} , \lambda_{2}}&=&\sum_{n=0}^{\infty} n P_{\lambda_{1}
      , \lambda_{2}}(n), \nonumber\\  _{\lambda_{1} , \lambda_{2}}\langle n^{2} \rangle_{\lambda_{1},
      \lambda_{2}}&=&\sum_{n=0}^{\infty} n^{2} P_{\lambda_{1} , \lambda_{2}}(n).
\end{eqnarray}
     To examine the statistics of the states, Mandel's $Q$-parameter is widely used, which characterizes the
     quantum statistics of the states of the field.
     This parameter has been defined as  $Q = {\langle n^2 \rangle - \langle n \rangle^2}/{\langle n \rangle} - 1$ \cite{Mandel}.
     To check Poissonian, sub-Poissonian or supper-Poissonian statistics,  as in \cite{drtvassoly1}, we can introduce a further extension definition for Mandel parameter \cite{Mandel} as follows:
\begin{eqnarray}\label{mh39}
     q_{\lambda_{1}, \lambda_{2}}=\frac{_{\lambda_{1} , \lambda_{2}}\langle n^{2} \rangle_{\lambda_{1} , \lambda_{2}}-
     _{\lambda_{1} , \lambda_{2}}\langle n \rangle_{\lambda_{1} , \lambda_{2}}^{2}}{_{\lambda_{1} , \lambda_{2}}\langle n
     \rangle_{\lambda_{1} , \lambda_{2}}}-1
\end{eqnarray}
     where $n=a^{\dagger}a$. \\

     We can also introduce an alternative deformed Mandel parameter, using the deformed number operator $N_{\lambda_{1},
     \lambda_{2}}=a^{\dagger}_{\lambda_{1} , \lambda_{2}}a$ as:
\begin{eqnarray}\label{mh41}
      Q_{\lambda_{1} , \lambda_{2}}=\frac{_{\lambda_{1} , \lambda_{2}}\langle N_{\lambda_{1} , \lambda_{2}}^{2}
      \rangle_{\lambda_{1}, \lambda_{2}} - _{\lambda_{1} , \lambda_{2}}\langle N_{\lambda_{1} , \lambda_{2}} \rangle_
      {\lambda_{1}, \lambda_{2}}^{2}}{_{\lambda_{1} , \lambda_{2}}\langle N_{\lambda_{1} , \lambda_{2}} \rangle_{\lambda_{1}, \lambda_{2}}}-1.
\end{eqnarray}
       When $Q_{\lambda_1 , \lambda_2}=0 $, the states exhibit Poissonian, $Q_{\lambda_{1} , \lambda_{2}}< 0 $
       sub-Poissonian, and $Q_{\lambda_{1} , \lambda_{2}}> 0 $ super-Poissonian statistics in $|n\rangle_{\lambda_{1},
       \lambda_{2}}$ basis. The same argument may be followed for $q_{\lambda_{1} , \lambda_{2}}$-parameter. It is interesting
       to notice that, while we have the lower bound $Q_{\lambda_{1} , \lambda_{2}}\geq -1$ there's no such boundary
       for $q_{\lambda_{1} , \lambda_{2}}$. Strictly speaking, $Q_{\lambda_{1} , \lambda_{2}}=0$ if the deformed number
       operator is considered, the situation that is the same as ordinary $Q$-parameter in orthonormal
       basis $|n\rangle$ when $n =a^{\dagger}a$.

\subsection{Quadrature squeezing}\label{mhl09}
       Quadrature squeezing is another property which the quantum states may possess. Firstly, we define $ x=\frac{1}
       {\sqrt{2}}(a+a^{\dagger})$, $p=\frac{1}{i\sqrt{2}}(a-a^{\dagger})$ and
       $\left( \Delta x_{i}\right) ^{2}= \langle x_{i}^{2} \rangle - \langle x_{i}\rangle^{2}$ where  $x_{i} = x , p.$
       Uncertainty relation in $x$ is obtained as
\begin{eqnarray}\label{mh48}
     (\Delta x ) ^{2}&=&  \frac{1}{2} [1+ (1- 2\lambda_{1})\langle a^{2} \rangle + \langle a^{\dagger 2} \rangle+2
     \langle a^{\dagger}_{\lambda_{1} , \lambda_{2}}a \rangle  \nonumber\\
     &-&2 \lambda_{2}\langle a \rangle - \langle a \rangle ^{2}
     - \langle a^{\dagger} \rangle ^{2} - 2\langle a \rangle \langle a^{\dagger} \rangle].
\end{eqnarray}
       Similarly for $p-$quadrature, one has
\begin{eqnarray}\label{mh50}
     (\Delta p) ^{2}&=&  \frac{1}{2} [1- (1+ 2\lambda_{1})\langle a^{2} \rangle - \langle a^{\dagger 2} \rangle +
      2\langle a^{\dagger}_{\lambda_{1} , \lambda_{2}}a \rangle  \nonumber\\
      &-& 2 \lambda_{2}\langle a \rangle + \langle a \rangle ^{2}
     + \langle a^{\dagger} \rangle^{2} - 2\langle a \rangle \langle a^{\dagger}
     \rangle],
\end{eqnarray}
      where all of the expectation values, should be calculated with respect to the $\lambda_1, \lambda_2$ coherent and squeezed states.
      \section{Numerical results and conclusion}\label{mhl11}
       From Figure 1 and 2, it is seen that the squeezing effect does not occur for the generalized CSs $|\alpha ,\lambda_{1},
       \lambda_{2}\rangle $. But as it is observed,  with increasing $\lambda_{1}$, for fixed $\lambda_{2}$-parameters the uncertainties in both quadratures
       tend to $0.5$ (the uncertainty of vacuum or canonical CSs).
       The same criterion for the SSs  $|\eta ,\lambda_{1},\lambda_{2}\rangle $ is investigated in Figure 3 and 4.
       The squeezing effect may be seen in $p$-quadrature for some fixed values of $\lambda_{1}$.

      In Figure 5 and 6 we displayed the Mandel parameter $Q_{\lambda_{1},\lambda_{2}}$ as a function of $\lambda_{1}$
       (for fixed values of $\lambda_{2}$) corresponding to respectively CSs $|\alpha ,\lambda_{1},\lambda_{2}\rangle$
       and SSs  $|\eta ,\lambda_{1},\lambda_{2}\rangle $. The negativity of this quantity in some regions, is observed
       from the two figures, showing the non-classicality of the introduced states. Figure 7 and 8 show the deformed
       Mandel parameter $q_{\lambda_{1},\lambda_{2}}$ defined in (\ref{mh39}) for again CSs $|\alpha ,\lambda_{1},\lambda_{2}\rangle$
       and SSs  $|\eta ,\lambda_{1},\lambda_{2}\rangle $, respectively. Negativity of this parameter in some regions of $\lambda_{2}$
       (for fixed values of $\lambda_{1}$) or in some regions of $\lambda_{1}$ (for fixed values of $\lambda_{2}$)  clearly shows the
       non-classicality features of these states.

          In conclusion, in this paper we have used the non-orthogonal squeezed states  \cite{xiang} $|n\rangle_{\lambda_{1} , \lambda_{2}}$
          which are the eigen-states of the non-Hermitian Hamiltonian $H_{\lambda_{1} , \lambda_{2}}$, as the bases for the construction of our CSs and SSs. We illustrated that, these states can be regarded as the bases of our infinite dimensional Hilbert space with a defined scalar product.
          Our motivation for this consideration is the more generality and more flexibility of the non-orthogonal
          basis: $\{|n\rangle_{\lambda_{1},\lambda_{2}} , n\in N\}_{n=0}^{\infty}$
          rather than orthogonal one $\{|n\rangle , n\in N\}_{n=0}^{\infty}$, and even the earlier one-parameter
          non-orthogonal bases $\{|n\rangle_{\lambda} , n\in N\}_{n=0}^{\infty}$, established by one of us \cite{drtvassoly1}.
          Also, the place of the deformations which lead to the outlined bases in the general framework of the representation theory
          of CSs is deeply established. Then, we concluded that, by some special superposition of the deformed Fock
          space, we can obtain new representations of CSs  $|\alpha ,\lambda_{1},\lambda_{2}\rangle $, as well as SSs
          $|\eta ,\lambda_{1},\lambda_{2}\rangle $ in the new two-parameters basis. Interestingly, in this Fock
          space, we obtained a set of new physical aspects; for instance, squeezing and sub-Poissonian statistics as some non-classical features.
          It is noticeable that, in the canonical CSs which is a
          composition of orthogonal bases, neither of these
          features may be observed. So, as it is implied in \cite{Ali2004}, in some classes of generalized CSs, which
          the non-classicality signs are revealed (nonclassical states), one may find
          their root in the non-orthogonality of basis,
          mathematically. Indeed, transforming the orthogonal
          basis to non-orthogonal one in the "canonical coherent states", results in the  appearance of
          the non-classicality features.


 {\bf Acknowledgement}
 The authors are grateful to Dr M Hatami and Dr M R Hooshmandasl  for their valuable helps in preparing the numerical results in final form.


\newpage

{\bf Figure Captions:}

=========================

{\bf Fig 1}
The graph of $(\triangle x)^{2}$ for CSs $\vert \alpha ,
        {\lambda_{1},\lambda_{2}}\rangle$,
     as a function of $\lambda_{1}$,  for $\lambda_{2}= 0.01, 2, 5.5, 8$, $ \alpha =0.8 $.

{\bf Fig 2}
 The graph of $(\triangle p)^{2}$ for CSs $\vert \alpha ,
   {\lambda_{1},\lambda_{2}}\rangle$,
   as a function of $\lambda_{1}$,  for $\lambda_{2}= 0.8,1.4, 2, 5 $, $ \alpha =0.8 $.

 {\bf Fig 3} The graph of $(\triangle x)^{2}$ for SSs $\vert \eta ,
   {\lambda_{1},\lambda_{2}}\rangle$,
   as a function of $\lambda_{2}$,  for $\lambda_{1}= -0.8,-0.4, 0.001, 0.6 $, $ \eta =0.8 $.

{\bf Fig 4}  The graph of $(\triangle p)^{2}$ for SSs $\vert \eta ,
    {\lambda_{1},\lambda_{2}}\rangle$,
    as a function of $\lambda_{2}$,  for $\lambda_{1}= -0.001, 0.5, 0.8 $, $ \eta =0.8 $.

{\bf Fig 5}  The graph of Mandel parameter $Q_{\lambda_{1},\lambda_{2}}$ for CSs $\vert \alpha,
    {\lambda_{1},\lambda_{2}}\rangle$ as a function of $\lambda_{1}$,  for $\lambda_{2} = -0.2, 0.0001, 0.5, 0.9$, $ \alpha =2 $.

{\bf Fig 6}   The graph of Mandel parameter $Q_{\lambda_{1},\lambda_{2}}$ for SSs $\vert \eta ,
    {\lambda_{1},\lambda_{2}}\rangle$ as a function of $\lambda_{2}$,  for $\lambda_{1} = -0.6, -0.3, 0.1, 0.8 $, $ \eta =0.8 $.

{\bf Fig 7} The graph of Mandel parameter $q_{\lambda_{1},\lambda_{2}}$  for CSs $\vert \alpha , {\lambda_{1},\lambda_{2}}\rangle$ as a function of
   $\lambda_{2}$,
   for $\lambda_{1} = -0.8, -0.1, 0.2, 0.8 $, $ \alpha =0.8 $.

{\bf Fig 8} The graph of Mandel parameter $q_{\lambda_{1},\lambda_{2}}$ for SSs $\vert \eta , {\lambda_{1},\lambda_{2}}\rangle$ as a function of
     $\lambda_{1}$,
      for $\lambda_{2}= -0.8, -0.2, 0.01, 0.6 $, $ \eta =0.8 $.

  \end{document}